# On the prevalence and scientific impact of duplicate publications in different scientific fields (1980-2007)

Vincent Larivière and Yves Gingras
Observatoire des sciences et des technologies (OST), Centre interuniversitaire de recherche sur la science et la technologie (CIRST), Université du Québec à Montréal, Montréal, Québec, (Canada)
Email: [lariviere.vincent; gingras.yves]@uqam.ca

## Abstract
The issue of duplicate publications has received a lot of attention in the medical literature, but much less in the information science community. This paper aims at analyzing the prevalence and scientific impact of duplicate publications across all fields of research between 1980 and 2007, using a definition of duplicate papers based on their metadata. It shows that in all fields combined, the prevalence of duplicates is one out of two-thousand papers, but is higher in the natural and medical sciences than in the social sciences and humanities. A very high proportion (>85%) of these papers are published the same year or one year apart, which suggest that most duplicate papers were submitted simultaneously. Furthermore, duplicate papers are generally published in journals with impact factors below the average of their field and obtain a lower number of citations. This paper provides clear evidence that the prevalence of duplicate papers is low and, more importantly, that the scientific impact of such papers is below average.

## Introduction
For obvious reasons of integrity and ethics in scientific publishing, the issue of duplicate publications has received a lot of attention, particularly in the medical literature. As Errami and Garner (2008) state, repeated publication of the same results, "not only artificially inflates an author's publication record but places an undue burden on journal editors and reviewers, and is expressly forbidden by most journal copyright rules." (p. 398). Duplicate publications can also affect the reliability of meta-analyses, as the same dataset might be counted more than once (Gurevitch and Hedges, 1999; Tramer *et al.*, 1997; Wood, 2008). Finally it is also in contradiction with Merton's norm of originality (Merton, 1973).

In order to avoid such publication behavior, the *New England Journal of Medicine* adopted, as early as 1969, a publication policy known as the Ingelfinger rule (Angell and Kassirer, 1991; Ingelfinger, 1969), which states that journals would only consider for publication manuscripts in which the substance has not been submitted or reported elsewhere. However, the debate continues, as several editorials aiming at creating guidelines and norms of publications have been recently published in medical journals such as the *Journal of Gastroenterology and Hepatology* (Farrell, 2007), the *Journal of Clinical Investigation* (Neill, 2008), *Research in Nursing & Health* (Baggs, 2008) as well as simultaneous editorials in journals *Surgery* and *British Journal of Surgery* (Murie *et al.*, 2006a,b). A group of medical researchers have also created a software that automatically finds duplicate papers as well as a public database that index them (Errami *et al.*, 2009). Contrary to the high level of interest among the medical community, the issue of duplicates has received much less attention in library and information science literature, although the field is not immunized from such practice. As shown by Davis (2005), several papers in the field of library and information science were actually exact duplicates published simultaneously by two journals. Between 1989 and 2003, Davis found 409 cases of republished papers.

Although the existence of duplicate papers raises important ethical questions, the goal of the present study is not to define an acceptable practice or provide moral condemnations or guidelines, but rather to contribute to the ongoing discussion by developing a bibliometric technique to produce macro-level data for all fields.

Before presenting our method to detect duplicate papers based on their metadata and analyze their characteristics, we will first survey the vast literature on the subject and recall the various methods used to identify them.

## Literature review

Previous macro-level studies on duplicates or highly related articles have mainly focused on the arXiv database (Sorokina *et al.*, 2006), which includes the full text of the papers, and on Medline, which indexes the metadata of the papers (Errami and Garner, 2008; Errami *et al.*, 2008; Errami *et al.*, 2009). Sorokina *et al.* (2006), comparing the full texts of the papers in ArXiv, found that 0.2%-0.5% of them were total or partial duplicates. Using advanced algorithms to compare titles and abstracts, Errami *et al.* (2008) have found that about 0.4% of Medline's entries were duplicates with different authors, and that 1.35% were duplicates with the same authors. Using the same algorithm (eTBLAST) on a larger subset of Medline (7 million articles and their most related articles—a function available on Medline), Errami and Garner (2008) found about 50,000 duplicates, representing about 0.7% of the database. This percentage has been increasing since the year 2000. These authors also found that about 20% of these duplicates were translations of the same article into another language.

At a more micro-level, several medical researchers have analyzed the prevalence of duplicates or highly related publications in their respective fields. Using a small sample of papers published in one journal in the field of otolaryngology, Rosenthal *et al.* (2003) found, by manually examining 492 papers, that about 8.5% of the papers were duplicates (same dataset and conclusions) or suspect duplicates (nearly identical dataset and conclusions). Using a similar qualitative and subjective method on larger samples of papers in the same field, Bailey (2002) found that 20% of authors had published an article with at least some degree of duplication, for a duplication rate of articles of about 1.8%. Bailey (2002) also showed that 4% of the duplicate papers were published in journals in another medical specialty. In ophthalmology, Mojon-Azzi *et al.* (2004) found that about 1.4% of the papers were duplicates, albeit 5% of them had their conclusions modified. Drawing on systematic reviews published in anesthesia and analgesia, von Elm *et al.* (2004) identified six different patterns of duplicate publication, ranging from papers with same sample of patients having the same outcomes to different sample of patients having different outcomes. They show that duplicate articles (that is the article published after the first one but identical to it) obtained lower citation rates and were published in journals with lower impact factors than the original article. Drawing on a small sample of papers in plastic surgery, Durani (2006) found, by examining the full text version of papers, that less than 1% of papers had some redundancy. Finally, Smith Blancett *et al.* (1995) showed, again by reading a small sample of papers in the field of nursing, that as high as 28% of the papers in nursing were either duplicates or segmented articles (salami slicing).

Given the many different methods used to define and thus identify duplicate papers, it is not surprising that their results differ greatly: duplicate articles can account for less than 1% to an astonishingly high of 28%—although most of the studies find the proportion of duplicate papers to be about 1%. The different definitions of what duplicate papers are reflect the many different conceptions of what ethical publication behavior is as well as a clear manifestation of the different publishing practices of different scientific fields.

## Methods

As discussed in the previous section, authors have used several definitions of what a duplicate paper is, which goes from one extreme (the exact same text is republished) to the other (e.g. only part of the data is reused, same data but different conclusions, etc.). The concept of a "highly-related paper" is highly subjective as intertextuality is inevitable and all papers must draw on previous research, including the authors' own previous publications. It also depends on what the authors consider ethical publishing. Finally, comparing the similarity of papers is not easy (is 50% or 75% the cutting line for being a duplicate paper?), we propose to define duplicates only by taking into account the metadata of the papers. This approach has the advantage that no subjective decision has to be taken and, more importantly, it is easily applicable to very large samples, such as all the papers contained in a given database. So we define as "duplicates" papers published in two different journals having the following metadata in common: 1) the exact same title, 2) the same first author, 3) the same number of cited references. The *number* of references was chosen instead of the references *themselves*, because a non-negligible proportion of references made to the same paper had the name of the author or of the journal written in a different manner. Of course, given that we have not read all the identified papers, we cannot exclude the possibility that these papers are in fact different despite these appearances. Although this possibility seems very remote given these three constraints, we nonetheless checked that possibility through an analysis of the duplicates for which the abstracts were available (N= 4,632) and found that more than 52% of the abstract were identical, and the remaining half very similar. Although this is a highly subjective task, the manual screening of a random sample of 100 of the remaining pairs of abstract (200 duplicates) showed that most of them (98%) were modified because of journals' different structure of abstracts or length or had only a few words added/removed and, thus, could not be considered to be "different" papers making different contributions to knowledge. We have also tested the other authors' order and names for 200 duplicates, and found differences in only 3.5% of the papers (author added, removed or changed in position)[1]. On this basis, we are confident that our method does identify real duplicate papers.

We use Thomson Reuters' Web of Science (WoS)—which includes the Science Citation Index Expanded, the Social Sciences Citation Index, and the Arts and Humanities Citation Index—to locate duplicate papers and measure their prevalence and scientific impact. In order to minimize the number of false positives, we excluded from the analysis papers published anonymously, those without any cited reference and with generic titles such as "Comment", "Commentary", "Untitled – Reply", etc. We also limited the analysis to research articles: all other document types ("review", "notes", "editorials", etc) were excluded because of the high number of false positives they generated. Articles labeled as "citation classics" were also excluded from the analysis, as well as document published in *Current Contents* (erroneously considered as a scientific journal in the WoS). Since the WoS does not index conference proceedings, conference papers subsequently published in scientific journals—a common practice in several disciplines—are not included in the study. Finally, papers republished more than twice (few in number) were excluded for reasons of homogeneity. Field and subfield classifications of journals are those used by the U.S. National Science Foundation (NSF). This classification categorizes each journal into one field and subfield. For the social sciences and humanities, the NSF categorization was complemented with our own classification of specialties.

[1] One might argue that identical papers are often published with different titles. Using the subset of all papers published in 2007, we have found only 34 occurrences (17 pairs) of papers with the same first author and list of references but different titles, which is quite lower than the number of papers found with our method. Hence, it seems that when researchers change their titles, they also change their lists of references. Using the same subset of data, we have also found 642 papers with the same title but different authors or number of references. 300 of these papers had both different authors and different references and 36 papers had the same number of references but different authors, which suggest that these are indeed different papers. A majority of these papers have short "generic" titles, making them more likely to be repeated. Finally, 306 papers had the same first author but a different number of references. These papers are more difficult to identify as duplicates or not: they have an identical title and, hence, are necessary related. Half of these papers had very similar abstract and could indeed be considered as "almost-identical" duplicates while the other half could be considered as highly-similar papers.

Given our strict definition of duplicate papers, our method necessarily excludes papers with the same content but with slight changes in the metadata. For example, duplicate papers for which the title, the references or the first author have been slightly modified are not considered as duplicates in this study. Obviously, papers not yet republished or republished in journals not indexed in the WoS are also excluded from the study. On the other hand, papers only published once but that have been indexed twice in the WoS—with a slight difference in the journal's name due to coding error—will be counted as duplicates, thus generating false positives. Our method will also count as duplicates unlikely cases where two really different papers share the same title, first author and number of references. For all these reasons, our method can be considered to provide a lower limit of duplicates and measure the tip of the iceberg; its complete size being contingent on researchers' opinions on what ethical publication behavior is. However, uniformly applied to all papers in the WoS database, our definition sheds light on the variable prevalence of duplicate papers in different scientific disciplines.

**Results**

Out of the 18,647,254 articles indexed in the WoS over the 1980-2007 period, we found 4,918 occurrences of duplicate papers published in two different journals, for a total of 9,836 entries. Based on our definition, duplicates thus account for 0.05% of the articles indexed in the WoS, which means that one paper out of two thousands has a duplicate published in another scientific journal. These duplicate articles received a total of 96,504 citations, for an average of slightly less than 10 citations per paper up to 2007. Figure 1 presents the annual number of duplicates and their percentage of the WoS database. The prevalence of such papers has been fluctuating over the period, with peaks in 1985 and 1999 (0.14% and 0.1% of the papers) and lows in 1995 and 2007 (0.02% and 0.03% of the papers), but the later may be due to duplicates to appear in 2008 or 2009. The absolute number of duplicates has nonetheless increased more or less steadily since the mid-1990s, which is consistent with the increase observed by Errami and Garner (2008).

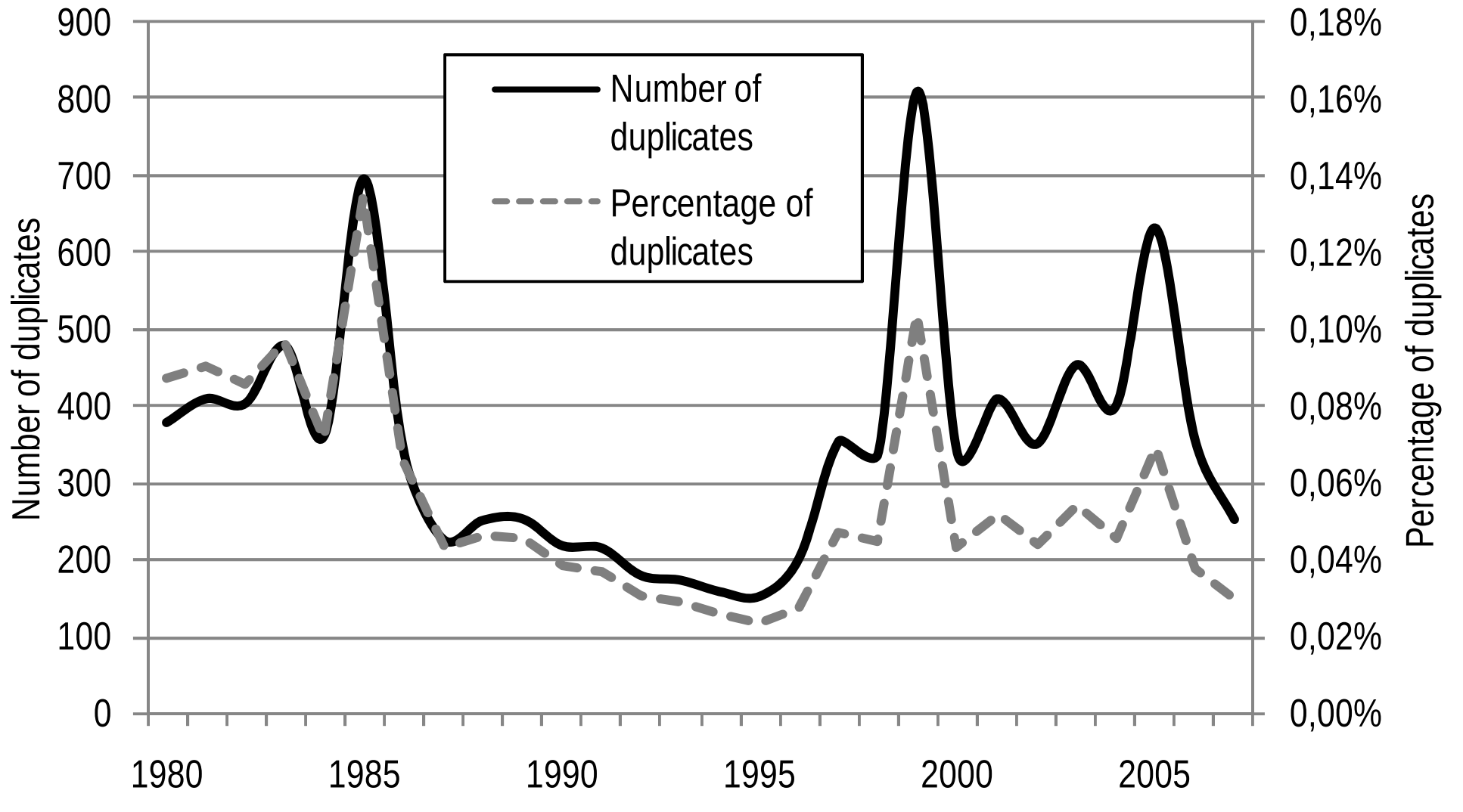

Figure 1. Number and percentage of duplicates, 1980-2007

Papers in languages other than English represent 12.3% of the dataset of duplicate papers, although they account for only 8% of the papers in the database. This percentage is lower than the 20% of non-English papers found by Errami and Garner (2008) but the difference could be due to the fact that they used

Medline and we use WoS. 8.2% of the duplicates are in a language other than English, 8.1% are duplicated in English and another language—likely translations from another language to English or vice-versa—and the remaining duplicates (86.7%) are both in English. Over the period studied, approximately three quarter of the duplicates were published in two journals in the same field; the remaining being published in journals in two different fields. This percentage is much higher than that of Bailey (2002), who only finds 4% of the duplicates published in a different specialty. This difference might be explained by the fact that Bailey only analyzed one field (otolaryngology–head and neck surgery), while our analysis includes all fields.

There is, however, much more traffic at the level of subfields. Indeed, only 55% of duplicates are published in the same subfield, which indicate that there is significantly more movement at a lower level of aggregation. Figure 2 is made with UCINET (Borgatti *et al.*, 2002) and NetDraw (Borgatti, 2002) softwares and present the relationship between the different subfields where duplicates are published. The size of the lines reflects the intensity of the relationship between the subfields, while the size of the nodes is a function of Freeman's (1979) degree of centrality. Only subfields-subfields relationships with 5 or more duplicates are shown. One can see that, in medical fields, a significant share of duplicate papers is published in a general medicine journal as well as into a more specialized journal. The same phenomenon applies to chemistry and physics, where the duplicate papers are published in a general and in a more specialized journal. There are also very strong links between applied fields such as applied physics, electrical engineering and electronics, and computers. This suggests that, although most duplicates are diffused in the same field, they target, in almost one case out of two, journals of different levels of specialization.

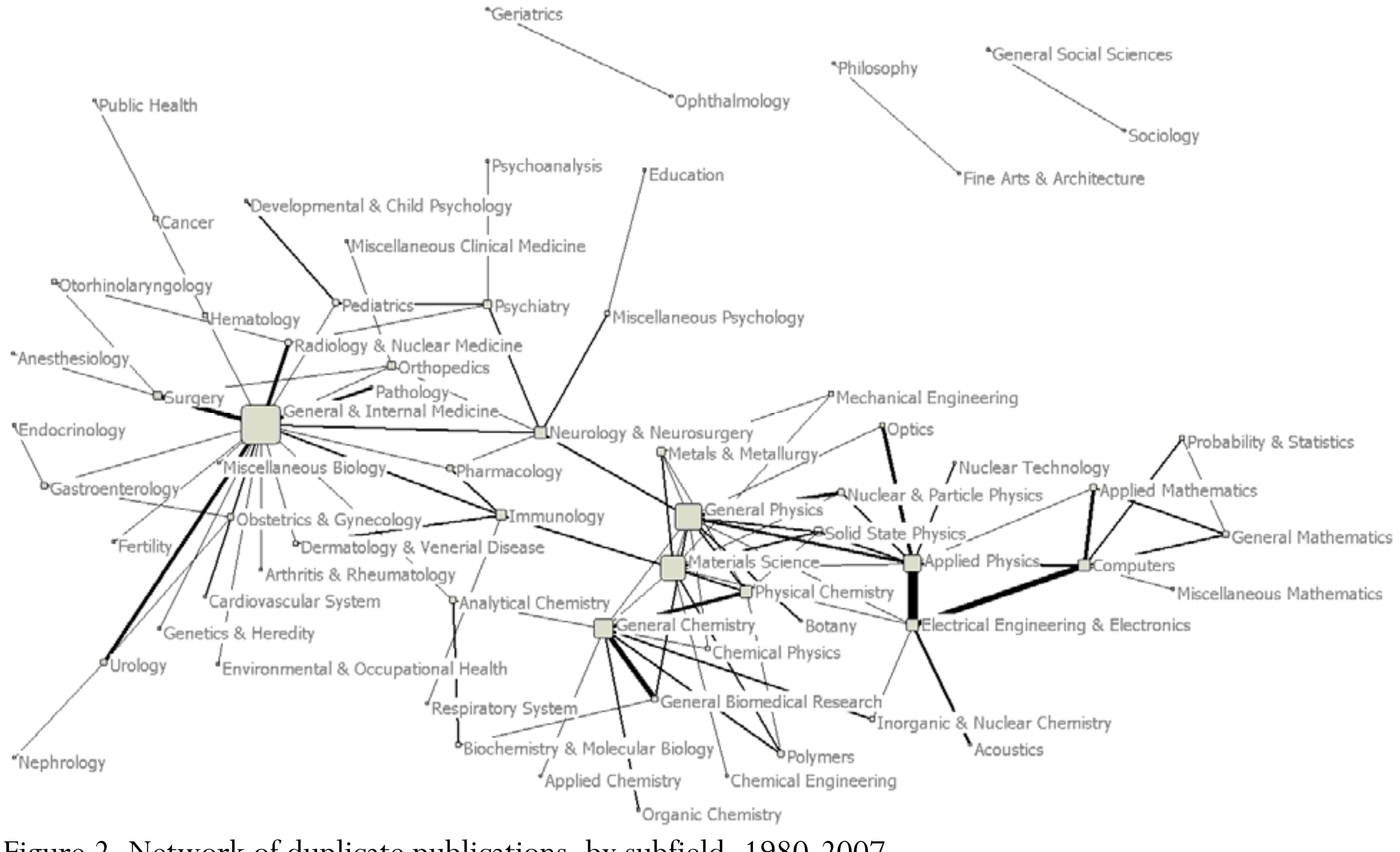

Figure 2. Network of duplicate publications, by subfield, 1980-2007.

Figure 3 shows that the proportion of duplicate papers varies greatly by field. Given the important number of studies on duplicate publications written in biomedical research and clinical medicine, one could have expected to find the highest proportion of duplicates among these fields. Surprisingly, that is not the case and biomedical research is rather among the lower end of the spectrum while clinical medicine is third after engineering and physics. Clinical Medicine is, however, the field with the higher absolute number of duplicates. The significantly higher proportion of duplicates found in engineering and technology can be explained at least in part by the fact that the WoS indexes several "proceedings" journals in these fields and that researchers might publish the same paper in both a proceeding (or a professional journal) and a more standard scientific journal. On the other hand, the fact that the arts and humanities have the lowest level of duplicates is likely related to their relying less on published papers than on books as an important outlet of diffusion of research than the social sciences which tend to conform more and more to the model of the natural sciences where papers has become the norm (Larivière *et al.* 2006). The low percentage of duplicates in arts and humanities is also, in all probability, a reflection of the low number of papers published in those fields.

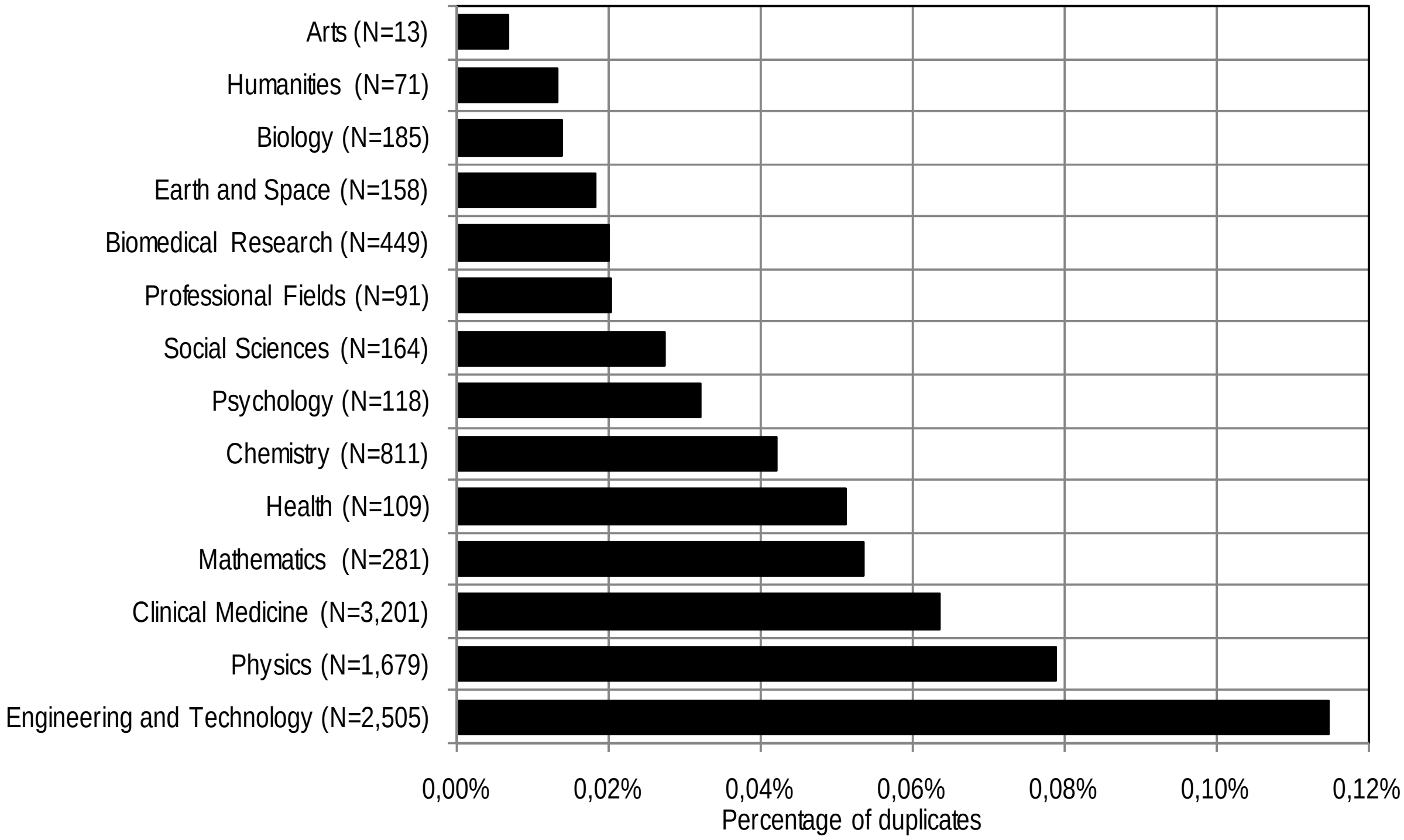

Figure 3. Percentage of duplicates, by field, 1980-2007

As shown in Figure 4, the average scientific impact of duplicate papers over the period studied—average of relative impact factor (ARIF) and average of relative citations (ARC)[2]—, is generally below the world average, a result consistent with those of Errami and Garner (2008), von Elm *et al.* (2004). Long *et al.* (2009) obtained the same results for plagiarized paper written by different authors. While impact measures of duplicates were above average in the 1980s, they dropped in the beginning of the 1990s and have remained below average since then. However, the trend varies according to the fields (Figure 5) though none are significantly more cited than the average of their field. In the natural and medical sciences duplicate papers are likely to be published in journals with lower impact factor and receive lower citations[3]. The fact that duplicate papers in social sciences tend to be more cited than is the case in the natural sciences may suggest that duplicates are less problematic in social sciences than in the physical and medical sciences as the very meaning of a "contribution to knowledge" maybe different and less tied to a specific "discovery" as understood for example in the physical sciences. Thus the publication of duplicates could even reflect the quality or demand of the paper, thus reprinted in different venues as it becomes better known.

[2] ARIF and ARC values above (or below) one means that the average scientific impact of paper is above (or below) the world average of their respective subfield. In order to take into account the fact that publication and citation practices vary according to fields, these scientific impact measures were normalized by the world average for each subfield (Schubert and Braun, 1986; Moed *et al.*, 1995). In the calculation of the impact factors, the asymmetry between the numerator and the denominator was also corrected. Indeed, in the calculation of their impact factors, Thomson Reuters counts citations received by all document types published (articles reviews, editorials, news items, etc.) but only divides these citations by the number of articles and reviews published, which are considered as "citable" items. This has the effect of artificially increasing the impact factor of journals with a higher ratio of "non–citable" items. For a historical review of impact factors' limits see Archambault and Larivière (2009).

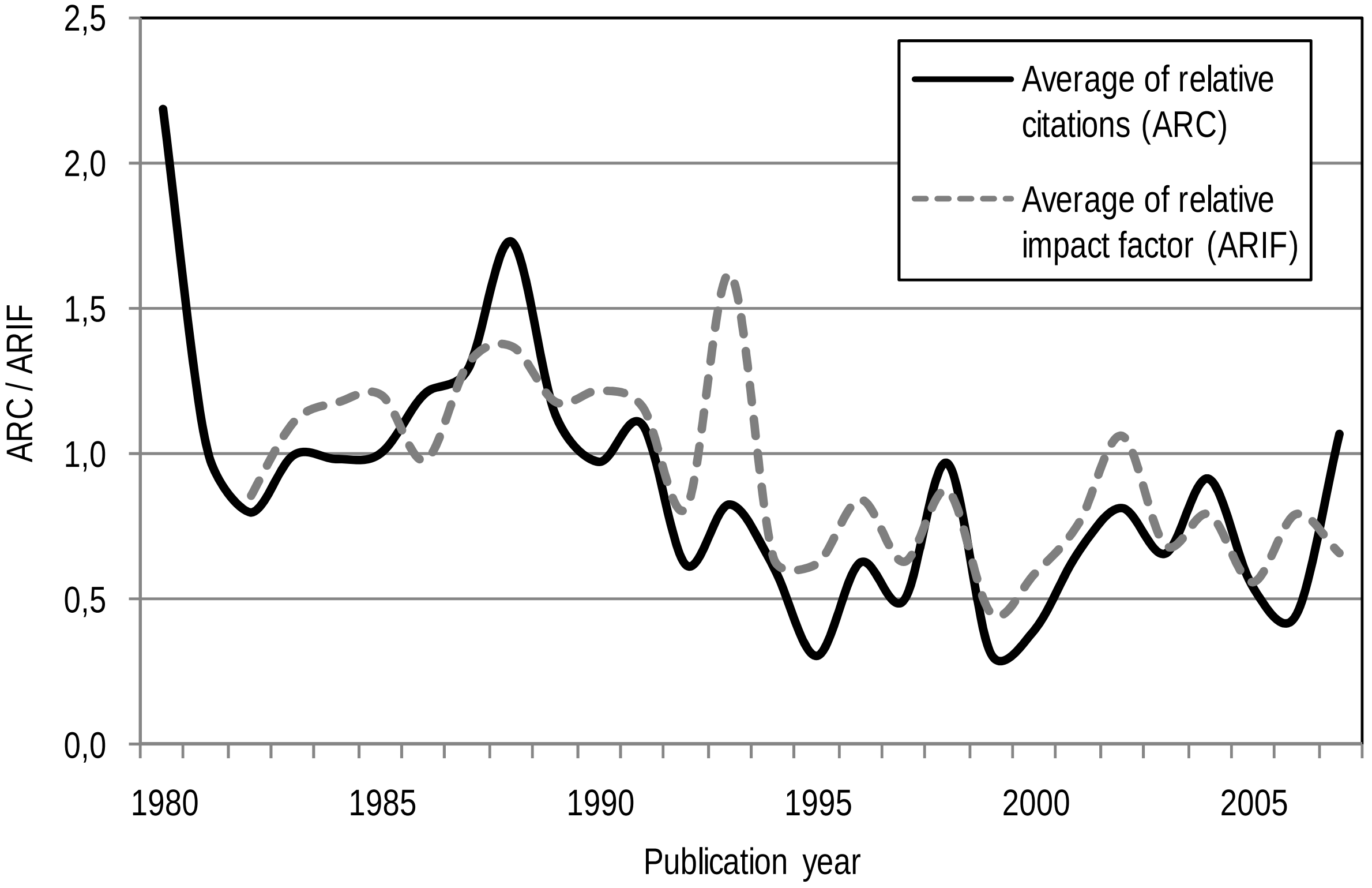

Figure 4. Average of relative citations and average of relative impact factor of duplicates, 1980-2007

---

[3] We have excluded fields with less than 100 papers, for which the fluctuations are large and not significant (arts, humanities and professional fields).

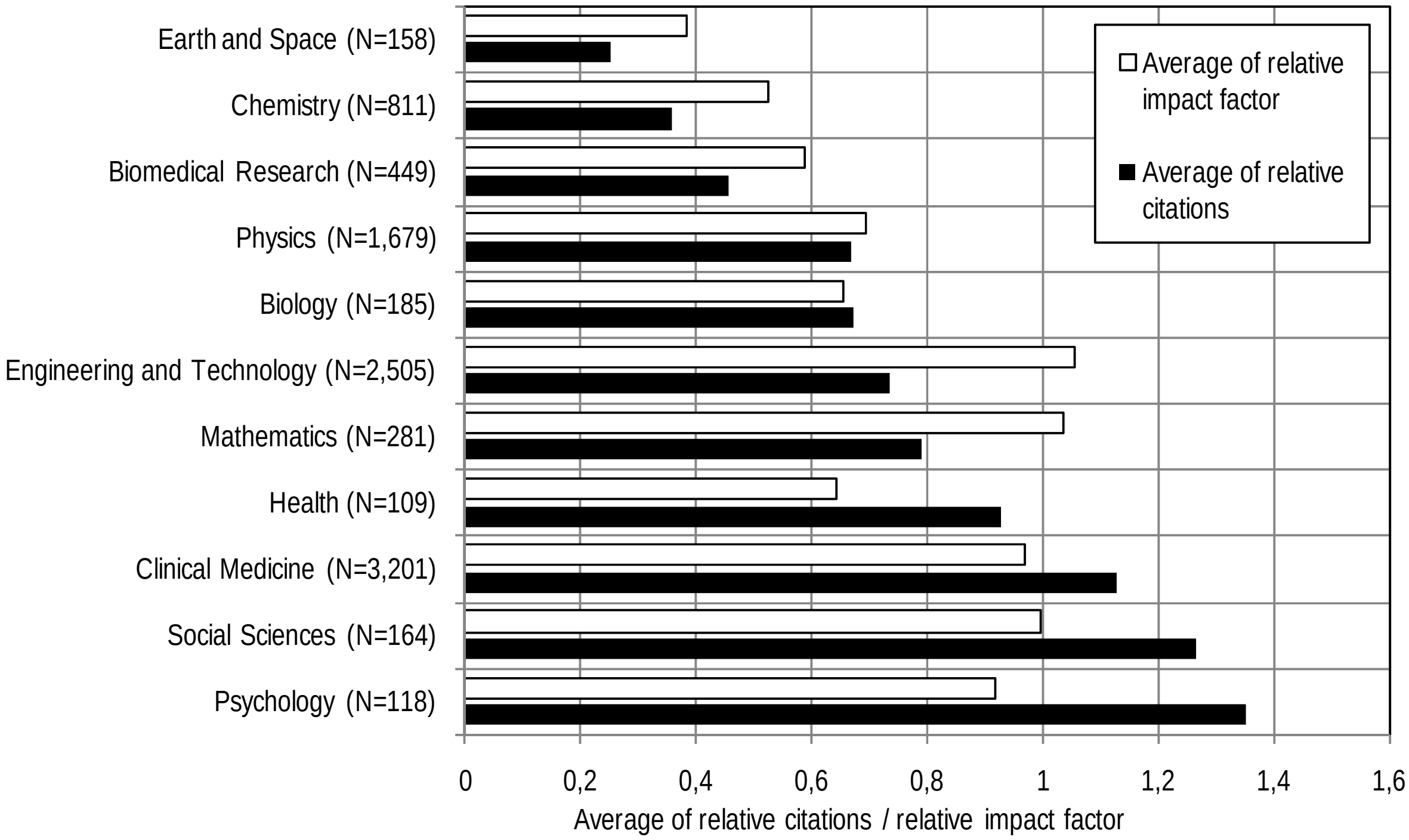

Figure 5. Average of relative citations and average of relative impact factor of duplicates, by field, 1980-2007

Most studies that have analyzed the difference between publication years of duplicates have shown that the majority of duplicate papers are published the same year (Errami and Garner, 2008; Rosenthal *et al.*, 2003). An exception is von Elm *et al*. (2004), who found more duplicates published in the year following the first publication. Our data (Figure 6) show that more than 60% of the duplicates are published during the same year and about 25% the subsequent year. This distribution suggests that the vast majority of these duplicates were submitted simultaneously to the two journals. As could be expected, there is a larger proportion of papers in Social sciences and humanities (23%) than in the natural sciences (12%) that are republished after 2 years or more. This fact again suggests that the meaning of republication is not the same for all disciplines and depends on the nature of the research contribution.

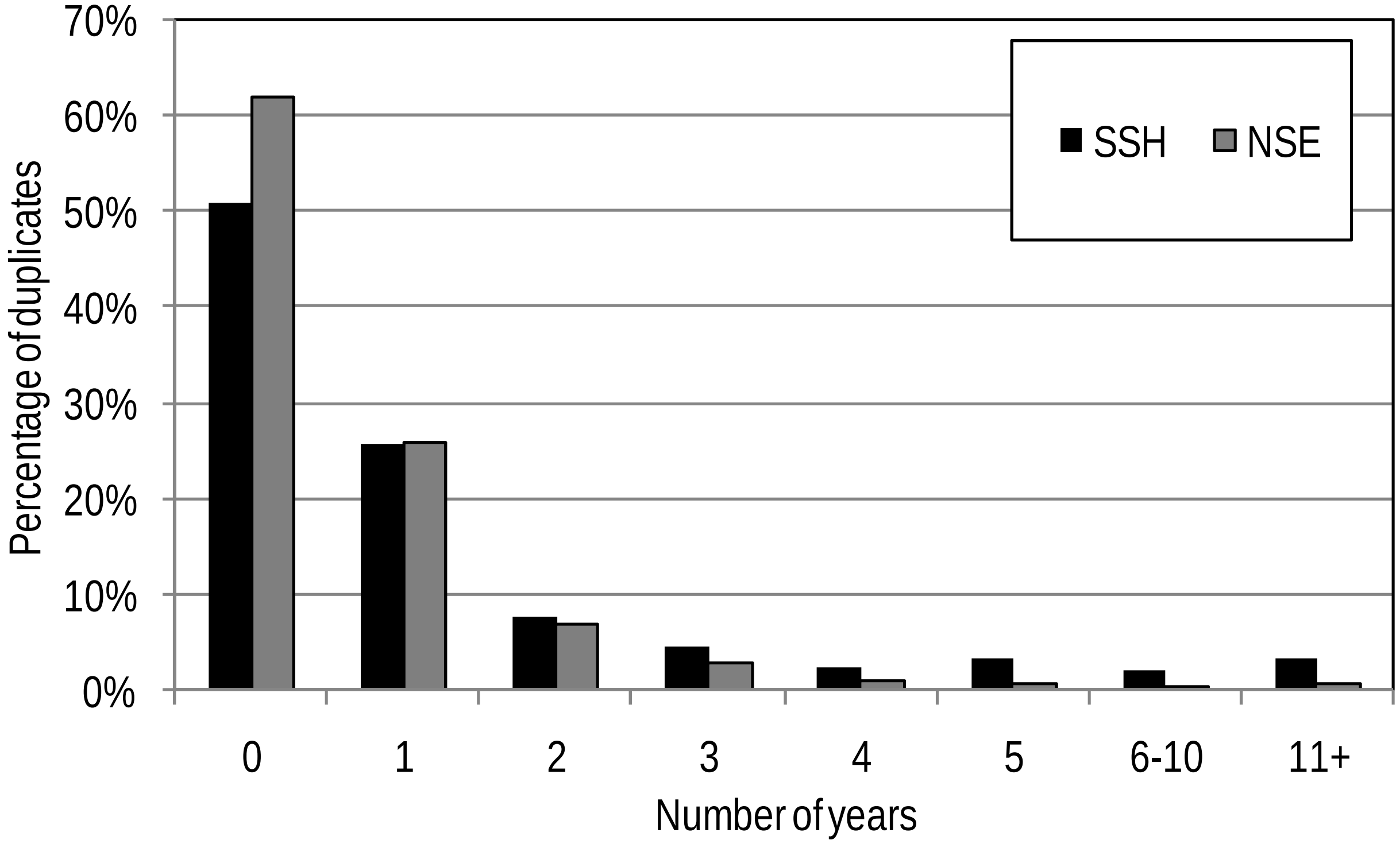

Figure 6. Percentage of duplicates, by number of years between the first and subsequent publication (three-year citation window), natural sciences and engineering (NSE) and social sciences and humanities (SSH), 1980-2007

Figure 7 presents, for the 3,820 duplicates published at different years, impact measures of both the paper first published and the subsequent paper. Unsurprisingly, the first publication generally receives higher citation counts, even with a fixed citation window of three years, a result which could be explained by a first mover advantage. However, it is more surprising to see that second papers published in year 2 and 3 after the first are often published in journals with higher relative impact factors. In fact, 55% of the papers published in the years following its first publication are published in journals higher relative impact factors. This seem to suggest that for some papers, the authors may have aimed too low in the hierarchy of journals and decided to republish the results in a better one or even have been invited to do so. But only a qualitative analysis based on interviews could clarify this point. Note, however, that this rise in relative impact factors is observed only in the natural sciences and engineering; there is no specific trends in social sciences and humanities.

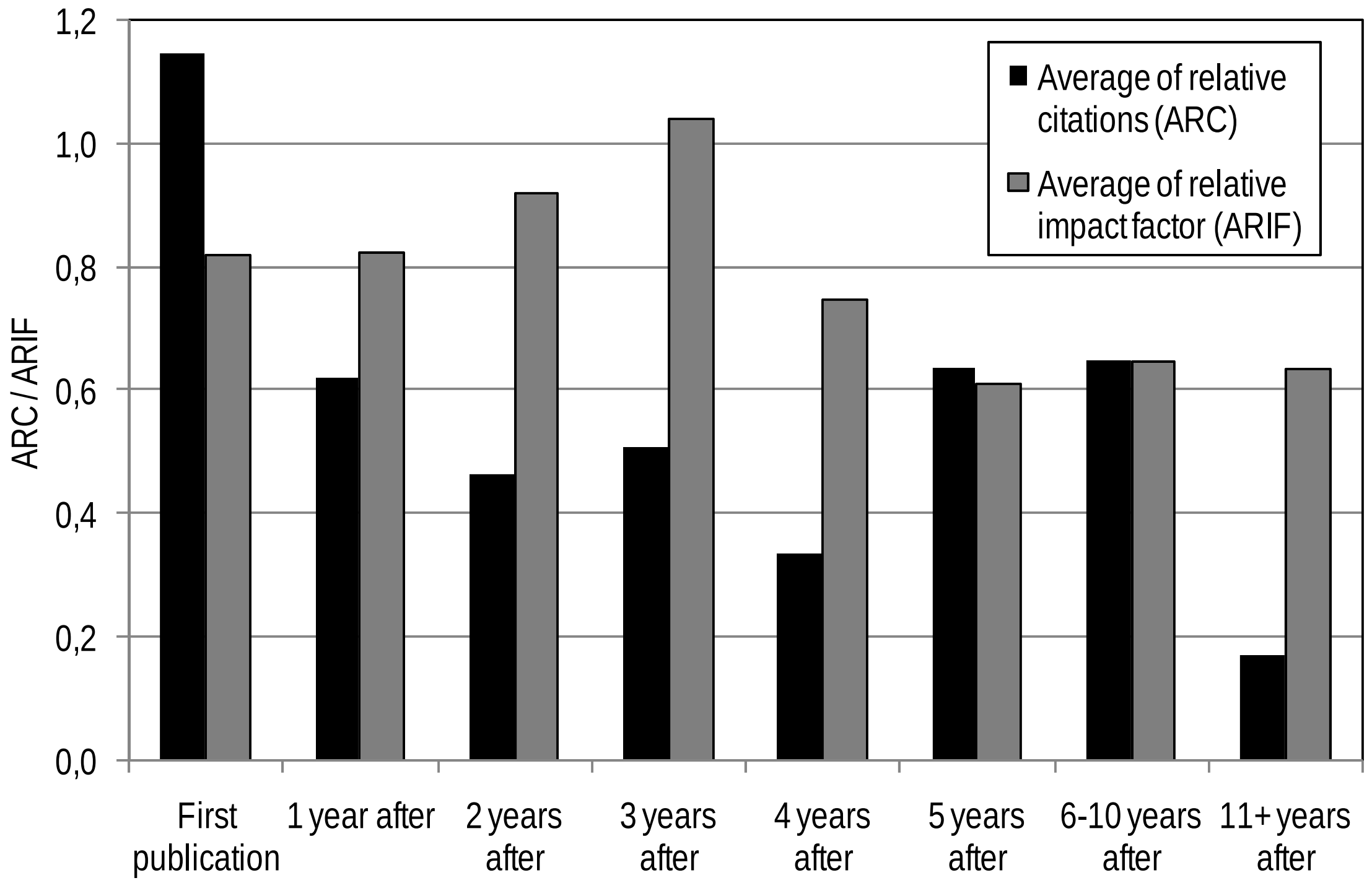

Figure 7. Average of relative citations and average of relative impact factor of duplicates published at two different years, 1980-2007

**Discussion and conclusion**

This study shows that the minimum prevalence of duplicates is about one paper per two thousands. Their numbers also seems on the rise since the mid-1990s, and, given our definition of duplicates, our results can be considered as a lower limit. Indeed, duplicate papers that slightly modify their metadata (titles, authors, references) cannot be captured by our method. Hence, the phenomenon of republishing papers—in whole or in part—or of segmenting papers is certainly more important than what our data suggest, though the percentage of 28% found by Smith Blancett *et al.* (1995) could be considered as the upper limit. Contrary to what could be expected, the prevalence of duplicate papers in the medical sciences is not higher than in physics and engineering and technology.

Although three out of four duplicates are published in journals of the same field, almost 50% of them are published in journals of two different subfields. More importantly, in a lot of cases, these papers are published in a general journal as well as in a more specialized journal. This is likely a reflection of the fact that in some fields, traditional scientific journals share the publication landscape with more "professional" journals, in which researchers might also diffuse their research results to reach different practitioners. One must nonetheless keep in mind that there are still more than 50% of the duplicate papers that are both published in the same subfield and, hence, that are not clearly aimed at reaching different potential audiences.

The rise of duplicate papers seems to be part of a context in which researchers are increasingly pushed to publish (or perish) in order to get tenure or continue to be funded while most disciplines are in a period of normal science (Kuhn, 1962) or in a steady-state (Ziman, 1994; Larivière *et al.*, 2008) where it is increasingly difficult to be truly original. But the case of the humanities and maybe to a lesser extent the social sciences, remind us that the notion of a "contribution to knowledge" may also affect the tendency to produce "duplicate" papers. For the physical and medical sciences however it is interesting to observe that such practices do not seem to be rewarded with more citations, that is with more scientific capital (Bourdieu, 2004).

**Acknowledgments**
The authors wish to thank Lorie Kloda, Jean Lebel and Benoit Macaluso for their useful comments and suggestions.

**References**
Angell, M. and Kassirer, J.P. (1991), "The Ingelfinger Rule revisited" *New England Journal of Medicine* Vol. 325 No. 19, pp. 1371-1373.

Archambault, E. and Larivière, V. (2009), "History of the journal impact factor: Contingencies and consequences", *Scientometrics*, Vol. 79 No. 3. In press.

Baggs, J.G. (2008), "Issues and rules for authors concerning authorship versus acknowledgements, dual publication, self plagiarism, and salami publishing", *Research in Nursing & Health*, Vol. 31 No. 4, pp. 295-297.

Bailey, B.J. (2002), "Duplicate publication in the field of otolaryngology–head and neck surgery", *Otolaryngology - Head and Neck Surgery*, Vol. 126 No. 3, pp. 211-216.

Borgatti, S.P. (2002), *NetDraw: Graph Visualization Software*, Analytic Technologies, Harvard, MA.

Borgatti, S.P., Everett, M.G. and Freeman, L.C. (2002), *Ucinet for Windows: Software for Social Network Analysis*, Analytic Technologies, Harvard, MA.

Bourdieu, P. (2004), *The Science of Science and Reflexivity,* University of Chicago Press, Chicago, IL.

Davis, P.M. (2005), "The ethics of republishing: A case study of Emerald/MCB university press journals", *Library Resources & Technical Services*, Vol. 49 No. 2, pp. 72-78.

Durani, P. (2006), "Duplicate publications: redundancy in plastic surgery literature", *Journal of Plastic, Reconstructive & Aesthetic Surgery*, Vol. 59, pp. 975-977.

von Elm, E., Poglia, G., Walder, B. and Tramèr, M.R.. (2004), "Different patterns of duplicate publication: An analysis of articles used in systematic reviews", *Journal of the American Medical Association*, Vol. 291(8), pp. 974-980.

Errami, M. and Garner, H. (2008), "A tale of two citations", *Nature*, Vol. 451 No. 7177, pp. 397-399.

Errami, M., Hicks, J.M., Fisher, W., Trusty, D., Wren, J.D., Long, T.C. and Garner, H.R. (2008), "Déjà vu—A study of duplicate citations in Medline", *Bioinformatics*, Vol. 24 No. 2, pp. 243-249.

Errami, M., Sun, Z., Long, T.C., George, A.C. and Garner, H.R. (2009), "Déjà vu: A database of highly similar citations in the scientific literature", *Nucleic Acids Research*, Vol. 37 Suppl. 1, pp. D921-D924.

Farrell, G.C. (2007), "Déjà vu, mais pas en anglais! Precautionary notes on publishing the same article in two languages", *Journal of Gastroenterology and Hepatology*, Vol. 22 No. 11, pp. 1699-1700.

Freeman, L.C. (1979), "Centrality in networks: I. Conceptual clarification", *Social Networks*, Vol. 1 No. 3, pp. 215-239.

Gurevitch, J. and Hedges, L.V. (1999), "Statistical issues in ecological meta-analysis", *Ecology*, Vol. 80 No. 4, pp. 1142-1149.

Huston, P. and Moher, D. (1996), "Redundancy, disaggregation, and the integrity of medical research", *Lancet*, Vol. 347 No. 9007, pp. 1024-1026.

Ingelfinger, F.J. (1969), "Definition of sole contribution", *New England Journal of Medicine*, Vol. 28, pp. 676-677.

Kuhn, T.S. (1962), *The Structure of Scientific Revolutions,* University of Chicago Press, Chicago, IL.

Larivière, V., Archambault, É. and Gingras, Y. (2008), "Long-term variations in the aging of scientific literature: From exponential growth to steady-state science (1900–2004)", *Journal of the American Society for Information Science and Technology*, Vol. 59 No. 11, pp. 288-296.

Larivière, V., Archambault, É., Gingras, Y. and Vignola-Gagné, É. (2006), "The place of serials in referencing practices: Comparing natural sciences and engineering with social sciences and humanities", *Journal of the American Society for Information Science and Technology*, Vol. 57 No. 8, pp. 997-1004.

Long, T.C., Errami, M., George, A.C., Sun, Z. and Garner, H.R. (2009), "Responding to possible plagiarism", *Science*, Vol. 323 No. 5919, pp. 1293-1294.

Merton, R.K. (1968), "The Matthew effect in science", *Science*, 159 No. 3810, pp. 56-63.

Merton, R.K. (1973), *The Sociology of Science: Theoretical and Empirical Investigations,* University of Chicago Press, Chicago, IL.

Moed, H.F., De Bruin, R.E. and van Leeuwen, Th.N. (1995), "New bibliometric tools for the assessment of national research performance: database description, overview of indicators and first applications", *Scientometrics*, Vol. 33 No. 3, pp. 381-422.

Mojon-Azzi, S.M., Jiang, X., Wagner, U. and Mojon, D.S. (2004), "Redundant publications in scientific ophthalmologic journal. The tip of the iceberg?" *Ophthalmology*, Vol. 111 No. 5, pp. 863-866.

Murie, J.A., Sarr, M.G. and Warshaw, A.L. (2006a), "A tale of three papers", *Surgery*, Vol. 140 No. 6, pp. 1063-1064.

Murie, J.A., Sarr, M.G. and Warshaw, A.L. (2006b), “A tale of three papers”, *British Journal of Surgery*, Vol. 93 No. 12, pp. 1560-1562.

Neill, U.S. (2008), “Publish or perish, but at what cost?” *The Journal of Clinical Investigation*, Vol. 118 No. 7, pp. 2368.

Rosenthal, E.L., Masdon, J.L., Buckman, C. and Hawn, M. (2003), “Duplicate publications in the otolaryngology literature”, *Laryngoscope*, Vol. 113, pp. 772-774.

Schubert, A. and Braun, T. (1986), “Relative indicators and relational charts for comparative assessment of publication output and citation impact”, *Scientometrics*, Vol. 9 No. 5-6, pp. 281-291.

Smith Blancett, S., Flanagin, A. and Young, R.K. (1995), “Duplicate publication in the nursing literature”, *Image: Journal of Nursing Scholarship*, Vol. 27 No. 1, pp. 51-56.

Sorokina, D., Gehrke, J., Warner S. and Ginsparg, P. (2006), “Plagiarism Detection in arXiv”. *Proceedings of the 6th IEEE International Conference on Data Mining (ICDM'06)*, pp. 1070-1075.

Tramer, M.R., Reynolds, D.J.M., Moore, R.A. and McQuay, H.J. (1997), “Impact of covert duplicate publication on meta-analysis: A case study”, *British Medical Journal*, Vol. 315 No. 7109, pp. 635-640.

Wood, J. (2008), “Methodology for dealing with duplicate study effects in a meta-analysis”, *Organizational Research Methods*, Vol. 11 No. 1, pp. 79-95.

Ziman, J.M. (1994), *Prometeus bound: Science in a dynamic steady state*, Cambridge University Press, Cambridge, U.K.